\documentclass[11pt]{article}
\usepackage{graphicx}
\usepackage{amsmath}
\usepackage{amssymb}
\usepackage{amsxtra}

\title{Correspondence between future-included and future-not-included theories}
\author{Keiichi Nagao\\Faculty of Education, Ibaraki University, Mito 310-8512 
Japan, \\
email: nagao@mx.ibaraki.ac.jp \\
and\\
Holger Bech Nielsen\\Niels Bohr Institute, University of Copenhagen, 
17 Blegdamsvej, \\
Copenhagen $\phi$, Denmark, \\
email: hbech@nbi.dk}

\begin{document}
\maketitle

\begin{abstract}

We briefly review the correspondence principle proposed in 
ref.\cite{Nagao:2012mj}, which claims that 
if we regard a matrix element defined in terms of the future state 
at time $T_B$ and the past state at time $T_A$ as an expectation value in the complex action theory whose path runs over not only past but also future,  
the expectation value at the present time $t$ 
of a future-included theory for large $T_B-t$ and large $t- T_A$ 
corresponds to 
that of a future-not-included theory with a proper inner product for large $t- T_A$.  
This correspondence principle suggests that the future-included theory is not excluded 
phenomenologically.  

\end{abstract}

\section{Introduction}\label{s:intro}

Recently complex action theory (CAT) 
has been studied\cite{Bled2006,own} with the expectation that 
the imaginary part of the action would give some falsifiable predictions. 
Indeed many suggestions have been made for Higgs mass\cite{Nielsen:2007mj}, 
quantum mechanical philosophy\cite{newer1,Vaxjo2009,newer2}, 
some fine-tuning problems\cite{Nielsen2010qq,degenerate}, 
black holes\cite{Nielsen2009hq}, 
De Broglie-Bohm particle and a cut-off in loop diagrams\cite{Bled2010B}. 
Also, integration contours 
in the complex plane\cite{Garcia:1996np}\cite{Guralnik:2007rx}, 
complex Langevin equations\cite{Pehlevan:2007eq} and 
complexified solution set\cite{Ferrante:2008yq} have been studied. 
Especially in ref.\cite{Bled2006} the authors studied 
a future-included theory, i.e. 
the theory including not only a past time but also a future time 
as an integration interval of time. 
They introduced a future state $| B (T_B) \rangle$ at the final time $T_B$ 
besides the ordinary past state $| A (T_A) \rangle$ 
at the initial time $T_A$, 
where $T_B$ and $T_A$ are set to be $\infty$ and $-\infty$ respectively. 
The states $| A (T_A) \rangle$ and $| B (T_B) \rangle$ 
time-develop according to the non-hermitian Hamiltonian $\hat{H}$ and $\hat{H}_B$, respectively, 
where $\hat{H}_B$ is set to be equal to $\hat{H}^\dag$\cite{Nagao:2011is}. 
They studied the matrix element\footnote{In the RAT the matrix element 
$\langle {\cal O} \rangle^{BA}$ is called weak value\cite{AAV} and 
has been intensively studied.} 
of some operator ${\cal O}$, 
$\langle {\cal O} \rangle^{BA} 
\equiv \frac{ \langle B(t) |  {\cal O}  | A(t) \rangle }{ \langle B(t) | A(t) \rangle }$, 
where $t$ is the present time, and 
speculated a correspondence of a future-included theory to 
a future-not-included one, i.e. 
$\langle {\cal O} \rangle^{BA} \simeq 
\langle {\cal O} \rangle^{AA} \equiv \frac{ \langle A(t) |  {\cal O}  | A(t) \rangle }{ \langle A(t) | A(t) \rangle }$.

In ref.\cite{Nagao:2012mj} 
we examined the quantity $\langle {\cal O} \rangle^{BA}$ carefully 
and found that if we regard it as an expectation value in a future-included theory, then we 
obtain the Heisenberg equation, the Ehrenfest's theorem and a conserved probability current density. 
This result strongly suggests that we can regard $\langle {\cal O} \rangle^{BA}$ 
as the expectation value in the future-included theory, 
though it is a matrix element in a usual sense. 
Furthermore improving the argument in ref.\cite{Bled2006} on 
the correspondence of a future-included theory to a future-not-included one 
by using both the complex coordinate formalism\cite{Nagao:2011za} 
and the automatic hermiticity mechanism\cite{Nagao:2010xu,Nagao:2011za}, 
i.e., a mechanism 
for suppressing the anti-hermitian part of the Hamiltonian 
after a long time development in a system defined with a 
non-hermitian Hamiltonian\footnote{The Hamiltonian is generically non-hermitian, 
so it does not belong to a class of the PT symmetric non-hermitian Hamiltonians 
intensively studied recently\cite{PTsym_Hamiltonians,Geyer,Mostafazadeh_CPT_ip_2002}.}, 
we have obtained a correspondence principle that 
$\langle {\cal O} \rangle^{BA}$ for large $T_B-t$ and large $t-T_A$ is almost 
equivalent to $\langle {\cal O} \rangle_{Q'}^{AA}$ for large $t-T_A$, 
where $Q'$ is a hermitian operator which is used to define a proper 
inner product\footnote{Similar inner products were studied 
also in refs.\cite{Geyer,Mostafazadeh_CPT_ip_2002}.}. 
In this article, for simplicity without 
using the complex coordinate formalism by considering the real $q$ case, 
we briefly review the argument to obtain 
the correspondence principle proposed in ref.\cite{Nagao:2012mj}.


This paper is organized as follows. 
In section 2 we explain the future-included theory and give the definitions of 
the states $| A(t) \rangle$ and $| B(t) \rangle$. 
In section 3 we review the proper inner products 
for the Hamiltonians $\hat{H}$ and $\hat{H}_B=\hat{H}^\dag$,  
and the automatic hermiticity mechanism. 
In section 4 we show that the expectation value of 
the future-included theory for large $T_B -t$ and large $t-T_A$ 
corresponds to that of  
the future-not-included theory with a proper inner product for large $t-T_A$. 
Section 5 is devoted to summary and outlook.

\section{Future-included theory}

A usual quantum theory is described with a real action and 
includes time integration from the past time to the present time. 
On the other hand we may be able to extend such a quantum theory 
so that it is described with a complex action and includes 
time integration from the past to the future. 
This is a future-included complex action theory (CAT), which we study 
in this article.

The future-included theory is described by introducing not only the 
ordinary past state $| A (T_A) \rangle$ 
at the initial time $T_A$ but also 
a future state $| B (T_B) \rangle$ at the final time $T_B$, 
where $T_A$ and $T_B$ are set to be $-\infty$ and $\infty$ respectively. 
In ref.\cite{Bled2006} 
the state $| A(t) \rangle$ and the other state $| B(t) \rangle$ at the present time 
$t$ 
are introduced\footnote{In ref.\cite{Nagao:2012mj} we have given the two slightly improved wave functions 
$\psi_A(q) = {}_m\langle_{new}~ q | A(t) \rangle$ and 
$\psi_B(q) = {}_m\langle_{new}~ q | B(t) \rangle$ 
based on the complex coordinate formalism\cite{Nagao:2011za} 
so that they are properly defined even for complex $q$. 
But in this article we consider only real $q$ case for 
simplicity and do not use the complex coordinate formalism.} by 
\begin{eqnarray}
&&\langle q | A(t) \rangle = \int_{\text{path}(t)=q} e^{\frac{i}{\hbar} S_{T_A ~\text{to} ~t}} D\text{path} , \\
&&\langle B(t) | q \rangle \equiv \int_{\text{path}(t)=q} e^{\frac{i}{\hbar} S_{t ~\text{to} ~T_B}} D\text{path} ,
\end{eqnarray}
where $\text{path}(t)=q$ means the boundary condition at the time $t$. 
The states $| A(t) \rangle$ and $| B(t) \rangle$ time-develop according to 
\begin{eqnarray}
&&i \hbar \frac{d}{dt} | A(t) \rangle = \hat{H} | A(t) \rangle , \label{schro_eq_Astate} \\
&&i \hbar \frac{d}{dt} | B(t) \rangle = \hat{H}_B | B(t) \rangle , \label{schro_eq_Bstate} 
\end{eqnarray}
where 
\begin{equation}
\hat{H}_B= \hat{H}^\dag . \label{defHB}
\end{equation}
We note that we explicitly derived the forms of $\hat{H}$ and $\hat{H}_B$
- for simplicity in a system with a single degree of freedom - 
via Feynman path integral in refs.\cite{Nagao:2011is}\cite{Nagao:2012mj} respectively. 
The authors in ref.\cite{Bled2006} speculated that the quantity 
\begin{equation}
\langle {\cal O} \rangle^{BA} 
=\frac{ \langle B(t) |  {\cal O}  | A(t) \rangle }{ \langle B(t) | A(t) \rangle } 
\label{expvalOBA}
\end{equation}
corresponds to 
\begin{equation}
\langle {\cal O} \rangle^{AA} 
=\frac{ \langle A(t) |  {\cal O}  | A(t) \rangle }{ \langle A(t) | A(t) \rangle } \label{OAA}
\end{equation}
in some approximation, i.e. 
\begin{equation}
\langle {\cal O} \rangle^{BA} \simeq \langle {\cal O} \rangle^{AA} . \label{expvalOBAOAA}
\end{equation}
The right-handed side is just an expectation value of $O$ in a usual future-not-included theory, 
while the left-hand side is not an expectation value but a matrix element of $O$ in a usual sense, and 
has the same form as the weak value\cite{AAV}.

\section{Proper inner products and the automatic hermiticity mechanism}

We briefly review the proper inner product for the Hamiltonians $\hat{H}$ 
and $\hat{H}_B$, and 
the automatic hermiticity mechanism\cite{Nagao:2010xu,Nagao:2011za}, 
i.e., a mechanism for suppressing the anti-hermitian part of the Hamiltonian 
after a long time development in a system defined with a non-hermitian Hamiltonian.

\subsection{A proper inner product for $\hat{H}$ and $\hat{H}_B$}

We introduce the eigenstates $| \lambda_i \rangle (i=1,2,\cdots)$ of 
the Hamiltonian $\hat{H}$ obeying 
$\hat{H} | \lambda_i \rangle = \lambda_i | \lambda_i \rangle$, 
where $\lambda_i (i=1,2,\cdots)$ are the eigenvalues of $\hat{H}$, 
and define the diagonalizing operator $P$ of $\hat{H}$ by 
$P=(| \lambda_1 \rangle , | \lambda_2 \rangle , \ldots)$. 
Then $\hat{H}$ is diagonalized as 
$\hat{H} = PD P^{-1}$, 
where $D$ is given by $\text{diag}(\lambda_1, \lambda_2, \cdots)$. 
Next we introduce an orthonormal basis $| e_i \rangle (i=1, \ldots)$ satisfying 
$\langle e_i | e_j \rangle = \delta_{ij}$ by 
$D | e_i \rangle = \lambda_i   | e_i \rangle$. 
The basis $| e_i \rangle$ is related to $| \lambda_i \rangle$ as 
$| \lambda_i \rangle = P | e_i \rangle$. 
Since $| \lambda_i \rangle$'s are not orthogonal to each other 
in the usual inner product $I$, 
$I(| \lambda_i \rangle , | \lambda_j \rangle ) = \langle \lambda_i | \lambda_j \rangle \neq \delta_{ij}$, 
the theory defined with $I$ would measure unphysical transitions. 
To make a physically reasonable measurement, we introduce 
a proper inner product $I_Q$\cite{Nagao:2010xu, Nagao:2011za} 
for arbitrary kets $|u \rangle$ 
and $|v \rangle$ as
\begin{equation}
I_Q(|u \rangle , |v \rangle) = \langle u |_Q v \rangle 
= \langle u | Q | v \rangle ,  
\end{equation}
where $Q$ is a hermitian operator chosen as 
\begin{equation}
Q=(P^\dag)^{-1} P^{-1}  \label{def_QA}
\end{equation}
so that the eigenstates of $\hat{H}$ get orthogonal to each other with regard to $I_Q$, 
$I_Q( | \lambda_i \rangle , |\lambda_j \rangle) = \delta_{ij}$. 
With this $I_Q$ we can make a physically reasonable observation and 
have the orthogonality relation 
$\sum_i | \lambda_i \rangle \langle \lambda_i |_{Q} = 1$. 
We note that $I_Q$ is different from the CPT inner product defined 
in the PT symmetric Hamiltonian formalism\cite{PTsym_Hamiltonians}.

We define the $Q$-hermitian conjugate of some operator $A$ by 
$A^{\dag^Q} = Q^{-1} A^\dag Q$. 
This satisfies 
$\langle \psi_2 |_Q A | \psi_1 \rangle^* = \langle \psi_1 |_Q A^{\dag^Q} | \psi_2 \rangle$. 
We also define $\dag^Q$ for kets and bras as 
$| \lambda \rangle^{\dag^Q} \equiv \langle \lambda |_Q $ and 
$\left(\langle \lambda |_Q \right)^{\dag^Q} \equiv | \lambda \rangle$. 
When some operator $A$ satisfies $A^{\dag^Q} = A$, 
we call $A$ $Q$-hermitian.\footnote{Similar inner products 
were studied also in refs.\cite{Geyer,Mostafazadeh_CPT_ip_2002}.} 
Since 
\begin{equation}
``P^{\dag^Q}"
\equiv
\left(
 \begin{array}{c}
      \langle \lambda_1 |_Q     \\
      \langle \lambda_2 |_Q     \\
      \vdots 
 \end{array}
\right) = P^{-1},
\end{equation}
satisfies 
$``P^{\dag^Q}" \hat{H} P = D$ and $``P^{\dag^Q}" \hat{H}^{\dag^Q} P = D^{\dag}$, 
$\hat{H}$ is $Q$-normal, 
$[\hat{H}, \hat{H}^{\dag^Q} ] = P [D, D^\dag ] P^{-1} =0$.  
In other words the inner product $I_Q$ is defined so that 
$\hat{H}$ is normal with regard to it.

Since $\hat{H}_B$ satisfies 
\begin{equation}
\hat{H}_B | \lambda_j \rangle_B = \lambda_j^* | \lambda_j \rangle_B , \label{HBlambdatildeket}
\end{equation}
where we have introduced 
$| \lambda_j \rangle_B \equiv Q| \lambda_j \rangle$, 
the diagonalizing matrix of $\hat{H}_B$ is given by 
$P_B \equiv (| \lambda_1 \rangle_B , | \lambda_2 \rangle_B , \ldots) 
=Q P = (P^\dag)^{-1}$. 
We introduce a proper inner product $I_{Q_B}$ 
for arbitrary kets $|u \rangle$ and $|v \rangle$ as 
$I_{Q_B}(|u \rangle , |v \rangle) = \langle u |_{Q_B} v \rangle 
= \langle u | {Q_B} | v \rangle$, 
where $Q_B$ is a hermitian operator chosen as 
\begin{equation}
Q_B
= (P_B^\dag)^{-1} P_B^{-1} 
= Q^{-1}   \label{QB_rel_Q-1}
\end{equation}
in order that $| \lambda_j \rangle_B$ get orthogonal to each other with regard to $I_{Q_B}$.  
We define $\hat{H}_B^{\dag^{Q_B}}$ by 
\begin{equation}
\hat{H}_B^{\dag^{Q_B}} = Q_B^{-1}  \hat{H}_B^\dag Q_B , \label{HBdagQB}
\end{equation}
which obeys 
${}_B\langle \lambda_i |_{Q_B} \hat{H}_B^{\dag^{Q_B}} = {}_B\langle \lambda_i |_{Q_B} \lambda_i$.

For later convenience we decompose $\hat{H}$ as 
$\hat{H}=\hat{H}_{Qh} + \hat{H}_{Qa}$. 
Here $\hat{H}_{Qh}= \frac{\hat{H} + \hat{H}^{\dag^Q} }{2} = P D_R P^{-1}$ and 
$\hat{H}_{Qa} = \frac{\hat{H} - \hat{H}^{\dag^Q} }{2}= i P D_I P^{-1}$ are 
$Q$-hermitian and anti-$Q$-hermitian parts of $\hat{H}$ respectively, 
where we have introduced $D_R= \frac{D + D^\dag }{2}$ and 
$D_I= \frac{D - D^\dag }{2}$.

\subsection{The automatic hermiticity mechanism} \label{automatic}

Following refs.\cite{Nagao:2010xu, Nagao:2011za} 
we study a time development of some state $| \psi(t) \rangle$ obeying 
the Schr\"{o}dinger equation 
$i \hbar \frac{d}{dt} | \psi (t) \rangle = \hat{H} | \psi (t) \rangle$. 
Since $| \psi' (t) \rangle \equiv P^{-1} | \psi (t) \rangle = \sum_i a_i(t) | e_i \rangle $ 
obeys $i \hbar \frac{d}{dt} | \psi'(t) \rangle =D | \psi'(t) \rangle$, 
$| \psi(t) \rangle = \sum_i a_i(t) | \lambda_i \rangle$ is expressed as 
\begin{equation}
| \psi(t) \rangle 
= \sum_i a_i(t_0) 
e^{ \frac{1}{\hbar} \left( \text{Im} \lambda_i - i \text{Re} \lambda_i \right) (t-t_0)}       
| \lambda_i \rangle . \
\end{equation}
Based on the assumption that the anti-hermitian part of $\hat{H}$ is bounded from above, 
which is needed to avoid the FPI $=\int e^{\frac{i}{\hbar}S} {\cal D}path$ 
divergently meaningless,  
we can crudely imagine that some of $\text{Im} \lambda_i $ 
take the maximal value $B$. 
We denote the corresponding subset of $\{ i \}$ as $A$. 
Then, if a long time has passed, namely for large $t-t_0$, 
the states with $\text{Im} \lambda_i |_{i \in A}$ survive and contribute most in the sum. 
We introduce a diagonalized Hamiltonian $\tilde{D}_{R}$ as 
\begin{equation}
\langle e_i | \tilde{D}_{R} | e_j \rangle \equiv 
\left\{ 
 \begin{array}{cc}
      \langle e_i | D_R | e_j \rangle =\delta_{ij} \text{Re} \lambda_i  & \text{for} \quad i \in A , \\
      0 &\text{for} \quad i \not\in A , \\ 
 \end{array}
\right. \label{DRtilder}
\end{equation}
and define $\hat{H}_{\text{eff}} \equiv P \tilde{D}_{R} P^{-1}$, 
which is $Q$-hermitian, 
and satisfies $\hat{H}_{\text{eff}} | \lambda_i \rangle = \text{Re} \lambda_i | \lambda_i \rangle$. 
Also, we introduce $| \tilde\psi(t) \rangle \equiv \sum_{i \in A}  a_i(t) | \lambda_i \rangle $. 
Then $| \psi(t) \rangle$ is approximately estimated as 
\begin{eqnarray}
| \psi(t) \rangle 
&\simeq& e^{ \frac{1}{\hbar} B (t-t_0)} 
\sum_{i \in A}  a_i(t_0) e^{-\frac{i}{\hbar} {\text Re} \lambda_i (t-t_0)} | \lambda_i \rangle \nonumber\\
&=&e^{ \frac{1}{\hbar} B (t-t_0)}  e^{-\frac{i}{\hbar} \hat{H}_{\text{eff}} (t-t_0)} 
| \tilde\psi(t_0) \rangle 
= | \tilde\psi(t) \rangle . \label{psiprimetket}
\end{eqnarray}
Thus we have effectively obtained a $Q$-hermitian Hamiltonian $\hat{H}_{\text{eff}}$ 
after a long time development. 
Indeed the normalized state 
\begin{equation}
| \psi(t) \rangle_{N} 
\equiv \frac{1}{\sqrt{ \langle {\psi}(t) |_Q ~{\psi}(t) \rangle} } | {\psi}(t) \rangle 
\simeq \frac{1}{\sqrt{ \langle \tilde{\psi}(t) |_Q ~\tilde{\psi}(t) \rangle} } | \tilde{\psi}(t) \rangle 
\equiv | \tilde{\psi}(t) \rangle_{N}
\end{equation} 
obeys the Schr\"{o}dinger equation  
\begin{equation}
i\hbar \frac{\partial}{ \partial t} | \tilde\psi(t) \rangle_{N} = \hat{H}_{\text{eff}} | \tilde\psi(t) \rangle_{N}.
\end{equation}
As we have seen above, the non-hermitian Hamiltonian $\hat{H}$ has become 
a hermitian one $\hat{H}_{\text{eff}}$ automatically 
with the proper inner product $I_Q$ and a long time development.

\section{Our analysis of $\langle {\cal O} \rangle^{BA}$}

We write eq.(\ref{expvalOBA}) as 
\begin{equation}
\langle {\cal O} \rangle^{BA} 
=\frac{  \langle A(t) | B(t) \rangle\langle B(t) |  {\cal O}  | A(t) \rangle }
{ \langle A(t) | B(t) \rangle \langle B(t) | A(t) \rangle } , \label{expvalOBA2} 
\end{equation}
and analyze it carefully. 
Using the expanded expression $| B(T_B) \rangle = \sum_i b_i | \lambda_i \rangle_B$ 
we obtain 
\begin{eqnarray}
| B(t) \rangle \langle B(t) |  
&=&  
e^{-i \hat{H}_B (t - T_B)} | B(T_B) \rangle  
\langle B(T_B) |_{Q_B}  e^{i \hat{H}_B^{\dag^{Q_B}} (t - T_B) }  
Q_B^{-1} \nonumber \\ 
&=&
\sum_{i,j} b_i  b_j^* e^{i  \text{Re}(\lambda_j - \lambda_i) (t - T_B)} 
e^{ \text{Im}(\lambda_j + \lambda_i) (T_B - t)} 
| \lambda_i \rangle_B  ~{}_B \langle \lambda_j |  \nonumber \\ 
&\simeq&  
\frac{1}{2\Delta t} \int_{t-\Delta t}^{t+ \Delta t} |B(t) \rangle\langle B(t)| dt  \nonumber \\
&\simeq&  
\sum_i | b_i |^2  e^{ 2\text{Im}(\lambda_i) (T_B - t)} | \lambda_i \rangle_B  ~{}_B \langle \lambda_i |  
\nonumber \\
&\simeq& e^{ 2B (T_B - t)}  Q_2  \quad  \text{for large $T_B - t$} ,
\label{Q2_0}
\end{eqnarray}
where in the third line we have smeared the present time $t$ a little bit, 
and then since the off-diagonal elements wash to $0$, we are lead to 
the fourth line. 
In the last line we have used the automatic hermiticity mechanism  
for large $T_B - t$, and $Q_2$ is given by 
\begin{eqnarray}
Q_2
&=&\sum_{i \in A} | b_i |^2 | \lambda_i \rangle_B  ~{}_B \langle \lambda_i |  \nonumber \\
&=&
\sum_{i \in A} F(\hat{H}_{\text{eff}}^B) | \lambda_i \rangle_B  ~{}_B \langle \lambda_i | \nonumber \\ 
&=&
F(\hat{H}_{\text{eff}}^\dag) Q \quad \text{for the restricted subspace} , 
\label{Q2finalexpression}  
\end{eqnarray}
where in the second equality assuming that $\text{Re} \lambda_i$'s are not degenerate, 
we have interpreted $|b_i|^2$ as a function of $\text{Re} \lambda_i$, 
$|b_i|^2 = F(\text{Re} \lambda_i)$. 
Also, 
$\hat{H}_{\text{eff}}^B \equiv P_B \tilde{D}_{R} P_B^{-1} = \hat{H}_{\text{eff}}^\dag$ 
is $Q_B$-hermitian,  
and obeys 
$\hat{H}_{\text{eff}}^B | \lambda_i \rangle_B = \text{Re} \lambda_i | \lambda_i \rangle_B$. 
In the last equality we have utilized the relation 
$\sum_{i \in A} | \lambda_i \rangle \langle \lambda_i |_{Q}=1$ 
for the subspace restricted by the subgroup $A$. 
For large $t-T_A$, 
since we have $| A(t) \rangle \equiv \sum_i  a_i(t) | \lambda_i \rangle 
\simeq \sum_{i \in A}  a_i(t) | \lambda_i \rangle \equiv | \tilde A(t) \rangle $  
by the automatic hermiticity mechanism 
eq.(\ref{expvalOBA2}) is expressed as 
\begin{equation}
\langle {\cal O} \rangle^{BA} 
\simeq
\frac{  \langle \tilde{A} (t) |_{Q_2}  {\cal O}  | \tilde{A}(t) \rangle }
{ \langle \tilde{A}(t) |_{Q_2}  \tilde{A}(t) \rangle }  
=\langle {\cal O} \rangle_{Q_2}^{\tilde{A} \tilde{A}}  
\quad \text{for large $t-T_A$}. \label{result_o_BA}
\end{equation}

Next we point out that the operator 
$P' = P f(D)$, where $f(D)$ is some function of $D$, 
is another diagonalizing matrix of $\hat{H}$, because 
$P' D {P'}^{-1} = P D P^{-1} = \hat{H}$. 
So we can define another inner product with $Q' = ({P'}^\dag )^{-1} {P'}^{-1}$. 
Choosing the function $f$ such that 
$(P^\dag )^{-1} f(D D^\dag)^{-1} P^\dag = F(\hat{H}^\dag)$ and 
using the automatic hermiticity mechanism for large $t-T_A$,
we obtain 
$Q' =  F(\hat{H}^\dag) Q 
\simeq F(\hat{H}_{\text{eff}}^\dag) Q  
= Q_2$ for the restricted subspace.  
Then  
the expectation value with the proper inner product $I_{Q'}$ in a future-not-included theory, 
which is introduced in refs.\cite{Nagao:2010xu, Nagao:2011za},  
is expressed as 
\begin{eqnarray}
\langle {\cal O} \rangle_{Q'}^{AA} 
&=&
\frac{  \langle A(t) |_{Q'} {\cal O}  | A(t) \rangle }
{ \langle A(t) |_{Q'} A(t) \rangle } \nonumber \\
&\simeq&  
\frac{  \langle \tilde{A}(t) |_{Q_2} {\cal O}  | \tilde{A}(t) \rangle }
{ \langle \tilde{A}(t) |_{Q_2} \tilde{A}(t) \rangle } 
=\langle {\cal O} \rangle_{Q_2}^{\tilde{A} \tilde{A}} 
\quad \text{for large $t-T_A$} .   
\label{expvalinfuturenotwithlongtime} 
\end{eqnarray}
Comparing eq.(\ref{result_o_BA}) with eq.(\ref{expvalinfuturenotwithlongtime}), 
we obtain the following correspondence: 
\begin{equation}
\langle {\cal O} \rangle^{B A} ~\text{for large $T_B-t$ and large $t-T_A$}
\quad \simeq   \quad 
\langle {\cal O} \rangle_{Q'}^{AA} ~\text{for large $t-T_A$}. \label{correspondence}
\end{equation}
This relation means that 
the future-included theory for large $T_B-t$ and large $t-T_A$ is almost 
equivalent to the future-not-included theory with a proper inner product for large $t-T_A$, 
and thus suggests that 
the future-included theory is {\em not excluded} though it seems exotic.

\section{Summary and outlook}

In ref.\cite{Bled2006} a correspondence of a future-included 
complex action theory (CAT) to a future-not included one was speculated,  
$\langle {\cal O} \rangle^{BA} \simeq \langle {\cal O} \rangle^{AA}$, 
where $\langle {\cal O} \rangle^{BA}$ and 
$\langle {\cal O} \rangle^{AA}$ are given in eqs.(\ref{expvalOBA})(\ref{OAA}) 
respectively. 
In ref.\cite{Nagao:2012mj} we studied $\langle {\cal O} \rangle^{BA}$ 
with more care by using the complex coordinate formalism\cite{Nagao:2011za} 
and the automatic hermiticity mechanism\cite{Nagao:2010xu}, 
i.e., a mechanism 
for suppressing the anti-hermitian part of the Hamiltonian 
after a long time development in a system defined with a 
non-hermitian Hamiltonian, and obtained our correspondence principle that 
$\langle {\cal O} \rangle^{B A}$ for large $t-T_A$ and large $T_B-t$ 
$\simeq$ $\langle {\cal O} \rangle_{Q'}^{AA}$ for large $t-T_A$, 
where $T_A$, $T_B$ and $t$ are the past initial time, the future final time and the present time, 
respectively. $\langle {\cal O} \rangle_{Q'}^{AA}$ 
is given in eq.(\ref{expvalinfuturenotwithlongtime}) 
and the $Q'$ is a hermitian operator used to define the proper inner product.

In this article we briefly reviewed the argument to obtain 
the correspondence principle 
following ref.\cite{Nagao:2012mj} without using 
the complex coordinate formalism\cite{Nagao:2011za} 
by considering the real $q$ case for simplicity. 
We first defined the two states $\langle B(t) |$ and $| A(t) \rangle$ 
from their respective functional integrals over future and past 
following ref.\cite{Bled2006} in section 2. 
In section 3 we reviewed the proper inner product and 
the automatic hermiticity mechanism\cite{Nagao:2010xu}. 
In section 4 we derived the correspondence principle 
following ref.\cite{Nagao:2012mj}. 
Thus the future-included theory for large $T_B-t$ and 
large $t-T_A$ is almost 
equivalent to the future-not-included theory with the proper inner product 
for large $t-T_A$, 
so such a future-included theory is not excluded phenomenologically.

In the correspondence principle the hermitian operator $Q'$ is a priori non-local, 
but it should be local phenomenologically. 
So we hope to invent some mechanism for getting it effectively local. 
Also, the other analyses in ref.\cite{Nagao:2012mj} suggest that 
the future-included theory looks more elegant in functional integral formulation 
than the future-not-included theory.
We will study the future-included theory in more detail and hope to 
report some progress in the future.

\section*{Acknowledgements}
The work of K.N. was supported in part by 
Grant-in-Aid for Scientific Research (Nos.18740127 and 21740157) 
from the Ministry of Education, Culture, Sports, Science and Technology (MEXT, Japan). 
K.N. would like to thank the organizers and participants of the workshop Bled 2012 
for their kind hospitality and useful discussions. 
K.N. is also grateful to Klara Pavicic 
for her kind support and hospitality during my stay in Volosko 
after joining this workshop.  
H.B.N. is thankful to NBI for allowing him to work at the institute as emeritus and 
to Mitja Breskvar for support of his travel to Bled and thereby also to Volosko.



\begin{thebibliography}{99}



\bibitem{Nagao:2012mj} 
  K.~Nagao and H.~B.~Nielsen,
  arXiv:1205.3706 [quant-ph], to be published in Prog.\ Theor.\ Exp.\ Phys.









\bibitem{Bled2006}
H.~B.~Nielsen and M.~Ninomiya, 
the proceedings of Bled 2006 
-What Comes Beyond the Standard Models-, 
p.87-124, 
arXiv:hep-ph/0612250.






\bibitem{own}

  H.~B.~Nielsen and M.~Ninomiya, 
  arXiv:0802.2991 [physics.gen-ph];
%
%
  Int.\ J.\ Mod.\ Phys.\  A {\bf 23} (2008) 919;
%
  Prog.\ Theor.\ Phys.\  {\bf 116} (2007) 851.









\bibitem{Nielsen:2007mj}
H.~B.~Nielsen and M.~Ninomiya, 
the proceedings of Bled 2007 
-What Comes Beyond the Standard Models-, p.144-185(arXiv:0711.3080 [hep-ph]).






\bibitem{newer1}
  H.~B.~Nielsen and M.~Ninomiya,
  arXiv:0910.0359 [hep-ph]. 



\bibitem{Vaxjo2009}
H.~B.~Nielsen, 
in Proceedings of the conference QTFR-5, June 14-18, 2009. Vaxjo, Sweden, 
arXiv:0911.4005 [quant-ph].





\bibitem{newer2}
H.~B.~Nielsen and M.~Ninomiya, 
the proceedings of Bled 2010 -What Comes Beyond the Standard Models-, 
p.138-157(arXiv:1008.0464 [physics.gen-ph]). 
  





\bibitem{Nielsen2010qq}
H.~B.~Nielsen,
arXiv:1006.2455 [physic.gen-ph].


\bibitem{degenerate}
H.~B.~Nielsen and M.~Ninomiya,
arXiv:hep-th/0701018.





\bibitem{Nielsen2009hq}
  H.~B.~Nielsen,
arXiv:0911.3859 [gr-qc].




\bibitem{Bled2010B}
H.~B.~Nielsen, M.~S.~Mankoc~Borstnik, K.~Nagao and G.~Moultaka, 
%
the proceedings of Bled 2010 
-What Comes Beyond the Standard Models-, 
p.211-216(arXiv:1012.0224 [hep-ph]).




\bibitem{Garcia:1996np}
  S.~Garcia, Z.~Guralnik and G.~S.~Guralnik,
  arXiv:hep-th/9612079.



\bibitem{Guralnik:2007rx}
  G.~Guralnik and Z.~Guralnik,
  Annals Phys.\  {\bf 325}, 2486 (2010). 



\bibitem{Pehlevan:2007eq}
  C.~Pehlevan and G.~Guralnik,
  Nucl.\ Phys.\  B {\bf 811}, 519 (2009). 


\bibitem{Ferrante:2008yq}
  D.~D.~Ferrante and G.~S.~Guralnik,
  arXiv:0809.2778 [hep-th].
  









\bibitem{Nagao:2011is}
  K.~Nagao and H.~B.~Nielsen, 
   Int.\ J.\ Mod.\ Phys.\ A.27, 1250076 (2012). 




		
\bibitem{AAV}
	Y. Aharonov, D. Z. Albert, and L. Vaidman,
	Phys. Rev. Lett. 
	{\bf 60}, 1351 (1988).





\bibitem{Nagao:2011za}
  K.~Nagao and H.~B.~Nielsen,
Prog.\ Theor.\ Phys. {\bf 126} No. 6, 1021 (2011) 
[errata: Prog.\ Theor.\ Phys. {\bf 127} No. 6, 1131 (2012)]. 






\bibitem{Nagao:2010xu}
K.~Nagao and H.~B.~Nielsen,
Prog.\ Theor.\ Phys. {\bf 125} No. 3, 633 (2011).






\bibitem{PTsym_Hamiltonians}

  C.~M.~Bender and S.~Boettcher,
  Phys.\ Rev.\ Lett.\  {\bf 80}, 5243 (1998).

  C.~M.~Bender, S.~Boettcher and P.~Meisinger,
  J.\ Math.\ Phys.\  {\bf 40}, 2201 (1999).
%





%
\bibitem{Geyer}
F. G. Scholtz, H. B. Geyer and F. J. W. Hahne, Ann. Phys. 213 (1992) 74.






\bibitem{Mostafazadeh_CPT_ip_2002}
A. Mostafazadeh, 
J.\ Math.\ Phys.\ {\bf 43}, 3944 (2002);  
%
J.\ Math.\ Phys.\ {\bf 44}, 974 (2003).




\end{thebibliography}
\end{document}